# Strength, corrosion resistance and cellular response of interfaces in bioresorbable poly-lactic acid/Mg fiber composites for orthopedic applications

Wahaaj Ali[1,2], Mónica Echeverry-Rendón[1], Alexander Kopp[3], Carlos González[1,4] and Javier LLorca[1,4]

[1]IMDEA Materials, C/Eric Kandel 2, 28906 Getafe, Madrid, Spain
[2]Department of Material Science and Engineering, Universidad Carlos III de Madrid, Leganés, Madrid 28911, Spain
[3]Meotec GmbH, Philipsstr. 8, 52068 Aachen, Germany
[4]Department of Materials Science, Polytechnic University of Madrid, 28040 Madrid, Spain

## Abstract

The shear strength and the corrosion resistance of the fiber/matrix interface after immersion in simulated body fluid was studied in poly-lactic acid/Mg fiber composites. The shear strength of the interface was measured by means of push-out tests in thin slices of the composite perpendicular to the fibers. It was found that the interface strength dropped from 15.2 ± 1.4 MPa to 7.8 ± 3.7 MPa after the composite was immersed in simulated body fluid for 148 hours. The reduction of the interface strength was associated to the fast corrosion of the fibers as water diffused to the interface through the polymer. The expansion of the fibers due to the formation of corrosion products was enough to promote radial cracks in the polymer matrix which facilitate the ingress of water and the development of corrosion pitting in the fibers. Moreover, cell culture testing on the material showed that early degradation of the Mg fibers affected the proliferation of pre-osteoblasts near the Mg fibers due to the local changes in the environment produced by the fiber corrosion. Thus, surface modification of Mg fibers to delay degradation seems to be a critical point for further development of Mg/PLA composites for biomedical applications.

1. **Introduction**

Biodegradable materials, including natural and synthetic polymers as well as hydrolizable metals, are currently considered for the fabrication of implants, tissue scaffolds and medical devices [1-3]. Among biodegradable polymers, the poly α-esters group, such as poly-lactic acid (PLA), poly-glycolic acid (PLG) and their co-polymers, have been used in sutures, drug-delivery systems, vascular stents, and orthopedic implants. PLA used for orthopaedic applications exhibit long degradation times (> 1 year), much longer than healing times [4]. Nevertheless, the accumulation of acidic products during degradation does not always favor tissue integration and, moreover, their low mechanical properties limit their use in many orthopedic applications. Natural polymers based of proteins (such as silk, collagen and fibrin) and polysaccharides (starch, alginate, etc.) are less likely to generate foreign body reactions but their mechanical properties are even lower than synthetic polymers. On the contrary, biodegradable metals, such as Mg, Zn and Fe, present much better mechanical properties and the corrosion products can be easily eliminated and/or metabolised or assimilated by cells [5]. Mg is particularly interesting for orthopedic applications because its elastic modulus (~45 GPa) is similar to that of human bones (2-30 GPa) [6] and Mg alloys are currently used in different orthopedic implants in the form of screws, plates, nails and wires for orthopedic surgeries [7-8]. However, the rapid corrosion of Mg in physiological environments is still a challenge because it can lead to early mechanical failures [9] and to the fast generation of hydrogen, a by-product of Mg corrosion, which is trapped in the tissue around implant causing inflammation or necrosis [10-12].

Considering the advantages and drawbacks of both types of biodegradable materials, Mg/polymer composites are an appealing solution because metallic fibers can provide the mechanical strength while the polymeric matrix slows down Mg corrosion. Furthermore, the mechanical properties and degradation rate of the composite can be tailored from the volume fraction and spatial distribution of matrix and fibers in the composite and it has been indicated the acidic products generated during the degradation of PLA can be neutralized by the degradation of Mg particles [13]. Li et al. [14] pioneered the fabrication of PLA matrix composites unidirectionally reinforced with Mg fibers and, since then, the processing and mechanical properties (tensile and bending strength as well as impact resistance) of these composites have been reported in the literature [15-19]. These investigations have shown that the elastic modulus and tensile strength of Mg fiber/PLA composites can reach similar values to those reported for load-bearing bones (tibia, femur) [15] and that the mechanical properties can be increased by means of fiber coatings [17, 20]. Nevertheless, these investigations were focused in the macroscopic mechanical properties and detailed analyses of the degradation mechanisms at the fiber/matrix interface as well as of the evolution of the interface properties were not provided. This information is critical because it is well-established that the fiber/matrix interface properties play a key role on the strength and toughness of composites [21-22].

In this investigation, the interface shear strength of PLA/Mg fiber composites as a function of immersion time in phosphate buffer saline (PBS) solution at 37°C is determined by means of push-out tests, that have been extensively used to measure the interface properties in other composites [23-28]. In addition, the fiber degradation micromechanisms at the interface are analyzed by means of scanning electron microscopy (SEM) and their influence of matrix cracking is assessed by means of an analytical model of the elastic stresses built up in the matrix due to the formation of corrosion of the Mg fiber. Finally, the cell response close and distant from the interfacial region is studied by means of direct tests. Overall, this investigation provides the first detailed assessment of the mechanical and biological performance of PLA/Mg fiber interfaces. This information is critical to design optimum Mg fiber coatings to enhance the *in vitro* and *in vivo* behavior of PLA/Mg composites.

## 2. Material and Experimental Techniques

### 2.1. Material

Billets of approximately 51 mm diameter of WE43MEO Mg alloy were produced by casting, machining and heat-treatment processes and then converted into 6 mm diameter rods by indirect extrusion (Meotec GmbH. Aachen). The nominal composition of the alloy was 1.4 – 4.2 % Y, 2.5 - 3.5 % Nd, <1 % (Al, Fe, Cu, Ni, Mn, Zn, Zr) and balance Mg (in wt%). Afterwards, the rods were cold drawn to fine wires of 0.1 mm diameter (Meotec gmBH, Aachen and Fort Wayne Metals Research Products Corp., Indiana, USA). The fibers had a constant diameter with tolerance of ±0.001 mm which was confirmed with a micrometer. The fibers were cleaned by the manufacturer to a standard appropriate for biomedical applications and were used as received.

PLA pellets (3251D) with a molecular weight around 160000 g/mol were supplied by NatureWorks BV (The Netherlands). Composite laminae were prepared by transferring fibers from spool to a Teflon covered plate. The fibers were loaded in tension using a custom-designed winding mechanism. A PLA/chloroform solution was then poured onto the fibers, leading to a unidirectional fiber composite lamina after evaporation and drying of the solvent. In addition, PLA laminae of 300 mm x 300 mm x 0.2 mm were prepared by hot pressing of the pellets. Alternative PLA and fiber composite laminae were placed into a mold of 120 mm x 12 mm x 2 mm and consolidated by hot pressing at 185ºC, leading to fully dense PLA/Mg fiber composite plates with an average fiber volume fraction of 20%.

### 2.2 *In vitro* degradation tests

Composites samples of 5 mm x 12 mm x 2 mm were cut from the plates for the degradation tests. This cross section (12 mm x 2 mm) is representative for bone plates and other biomedical applications. The cross-sections created during cutting (where the Mg fibers were exposed) were sealed with a waterproof epoxy resin to ensure that degradation occurred through the PLA matrix and not by pipeline diffusion mechanisms along the fiber/matrix interface from the cross-sections where Mg fibers are exposed. The samples were immersed in PBS solution at 37°C in a sealed polypropylene container. The PBS volume to sample mass ratio was above 30 ml /g and the PBS solution was changed daily. The samples were removed after 27 h (~1 day) and 148 h (~6 days), dried and thin slices for the push-out tests were prepared as indicated below. Visual inspection showed that ends of the specimens (in which you could see the cross section of the fibers) were effectively sealed because rapid corrosion at this location would have peeled of epoxy. It is worth noting that the intended period of *in vitro* study was 14 days, but preparation of thin slices was impossible after this period due to the enhanced degradation of composite. The surfaces of the push-out test samples were observed in the SEM (Zeiss EVO MA15) at an accelerated voltage of 3~5 kV to determine the degradation micromechanisms and corrosion products were identified by energy-dispersive X-ray spectroscopy (EDAX Octane Plus).

## 2.3 Push-out tests

Thin slices perpendicular to the fibers were cut from the composite plates using a high precision wire cutting machine for the push-out tests. The slices were embedded in an epoxy resin (SpeciFix resin system, Struers, Denmark) for successive grinding and polishing steps to obtain a sample with thickness of 0.2 mm, twice the diameter of fibers, ensuring that both the faces of the slice were perpendicular to the fiber direction. The thickness of each slices was measured with a Vernier caliper and they were in the range 0.19 to 0.21 mm.

Push-out tests of the fibers were carried out in the thin slices that were placed on a metallic support with a central groove of ≈ 200 μm in width (Fig. 1). The fiber to be tested was positioned manually at the center of groove with the help of an optical microscope and loaded in compression using a Hysitron TI 950 nanoindenter (Minneapolis, MN, USA) with a flat diamond tip of 70 μm in diameter under displacement control at 50 nm/s. 20 fibers were pushed out on 5 different samples of the as-manufactured composites while 8 and 5 fibers were pushed from the composite plates immersed in PBS for 27 and 148 hours, respectively. Special care was taken to ensure that the distance between the centers of nearest neighbor fibers and the pushed-out fiber was -at least- 150 μm (1.5 times the fiber diameter) to avoid constraint effects during the push-out test [26-27].

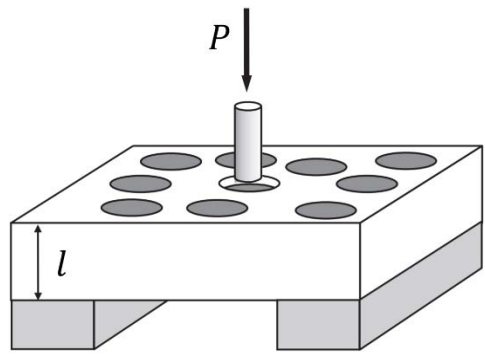

**Fig. 1.** Schematic of the fiber push-out test

The average interfacial shear strength between the Mg fiber and the PLA matrix was calculated according to [23-28]

$$\tau = \frac{P_{max}}{2\pi Rl} \qquad (1)$$

where $P_{max}$, $R$ and $l$ stand for the maximum load recorded during push-out test, the fiber radius and the thickness of composite slice, respectively.

## 2.3 Cell-composite interaction tests

A pre-osteoblast cell line from mouse calvarial MC3T3-E1 sub-clone 4 (ATCC CRL-2593) was used to study the interaction of the composite with cells. Cells were cultured in alpha-MEM medium (Invitrogen) supplemented with 10% fetal bovine serum (FBS) (Invitrogen), 2 mM glutamine, and 100 U/ml of penicillin and 100μg/ml streptomycin (Pen/Strep, Invitrogen). Cells between passages 11 and 13 were used and maintained at 37°C in a humidified 5% $CO_2$ atmosphere. The medium was renewed every two days, and cells were used after reaching 90% of confluence.

Composite samples with dimensions 0.5 cm x 1cm x 0.2 cm were sterilized by immersion in 70% ethanol during 10 min followed by air drying. Cells were directly seeded on the surfaces of the material at a concentration of 100.000 cell/cm$^2$ in 100 µl of alpha-MEM. After 30 min, 1 mL of complete culture medium was added and cells were incubated at 37°C for 24h. Finally, cells were washed twice with PBS and fixed with 4% formaldehyde solution for 30 min. Afterwards, they were passed through serial solutions of 30%, 50%, 70, 80%, 90%, and 100% ethanol each for 15 min. The samples were air-dried, sputter-coated with gold and observed in the SEM at an acceleration voltage of 7 kV.

## 3. Results

### 3.1 Interfacial behavior of Mg/PLA composite before degradation

Representative load-displacement curves of five push-out tests on isolated fibers of Mg/PLA composites are plotted in Fig. 2a. The initial shape of the curves was concave due to the elastic bending of the thin slice in the groove of the metallic support and was followed by a linear region with a slope of 200 ± 26 N/mm. The small differences in the slope between fibers could be attributed to the position of the fiber with respect to the groove. Further deformation led to a non-linear behavior which was associated with the progression of crack from the fiber/matrix interface at the bottom of the slice to the top. Other sources of non-linearity observed in push-out tests, such as plastic deformation of the matrix [26] or of the fibers [29] were not found during the visual inspection of the pushed-out fibers (Fig. 2b) and can be neglected due to the small load attained during the tests. The load dropped abruptly to zero after the maximum load was attained as a result of the complete fracture of the interface in shear, which was associated with push-out of the fiber (Figs. 2b and 2c). It should be noted that the final displacement of the fiber was much larger than the displacement of the indenter at the peak load, indicating a brittle failure mode.

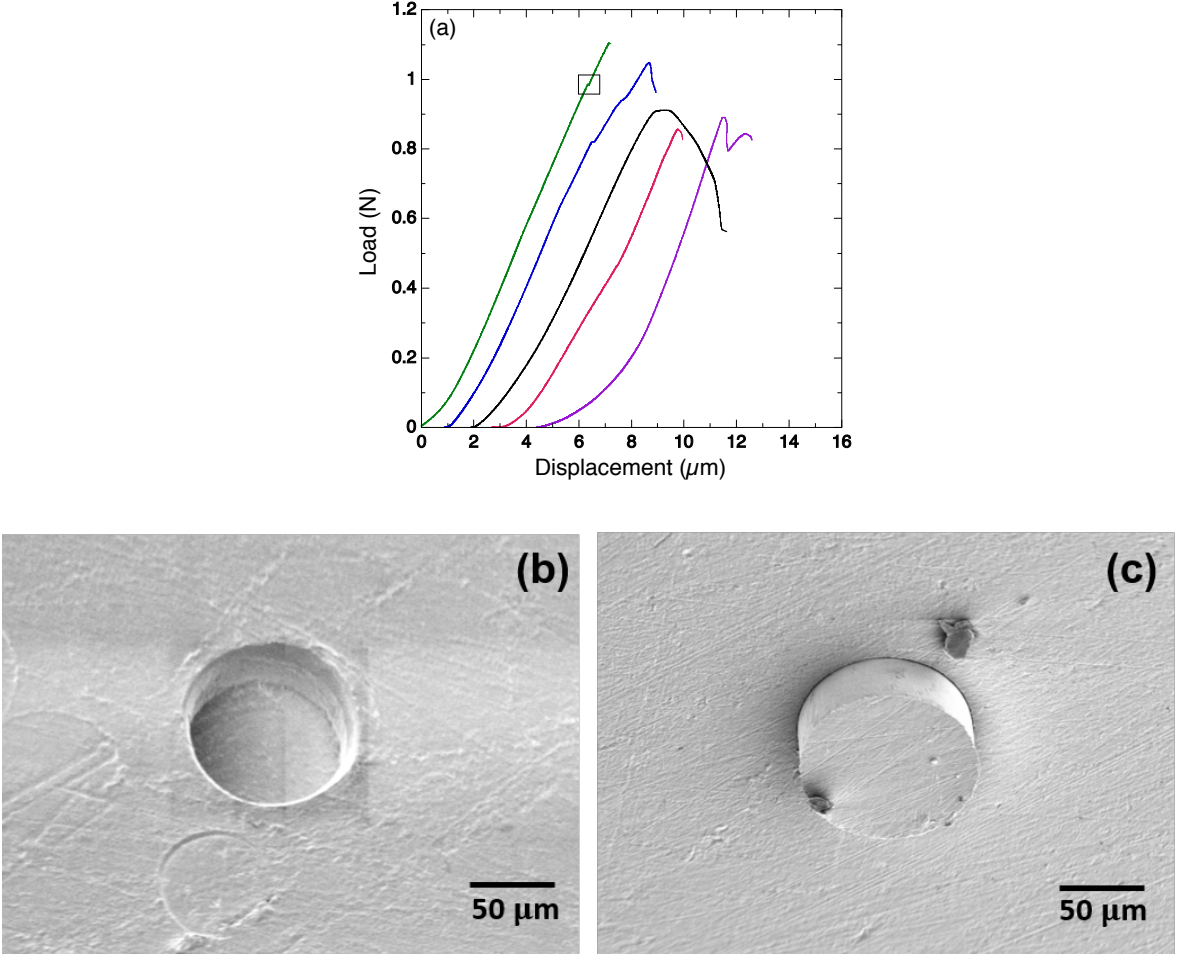

**Fig. 2.** (a) Representative load-displacement curves of the push-out tests in as-manufactured Mg/PLA composites. (b) Secondary electron SEM image of the top surface of the slice after the push-out test. (c) *Idem* of the bottom surface.

Some push-out curves showed a fluctuation in the load-displacement curve prior to the maximum load, as indicated by the square in Fig. 2a. It has been reported that these events are associated with the nucleation of small cracks at the interface at the bottom of the slice [22].

### 3.2 Interfacial behavior of Mg/PLA composite after degradation

The load-displacement curves of the push-out tests in composite plates immersed in PBS during 27 and 148 hours are plotted in Figs. 3a and 3b, respectively. The push-out tests in the samples immersed for 27 hours showed a more marked concave region at the beginning of the test (as compared with the as-manufactured samples), that could be attributed to uneven surfaces resulting from grinding/polishing because of the deterioration of the composite, indicating that the bending stiffness of the composite plate has decreased significantly. In addition, two samples showed an extended region with constant, negative slope after the peak-load, which was associated with the progressive push-out of the fiber against frictional forces at the broken interface. Nevertheless, interface failure was also brittle in these cases and the fibers protruded from the bottom of the slice at the end of the tests in all samples. The concave region at the beginning of the test and the extended region with negative slope after the peak load were more important in the samples tested after 148 hours in PBS (Fig. 3c). Moreover, the scatter in the shape of the load-displacement curves and in the peak-load increased dramatically due to the localized effects of corrosion. However, fiber pop-out from the bottom surfaces was never observed. Fluctuations in load-displacement curves before peak load were not observed in most of the degraded samples which is an indication of stable crack growth at interface. It should be noted, however, that the large deflection of the samples degraded during 148 h may lead to large normal tensile stresses at interface which could influence the actual value of the interface strength calculated from eq. (1).

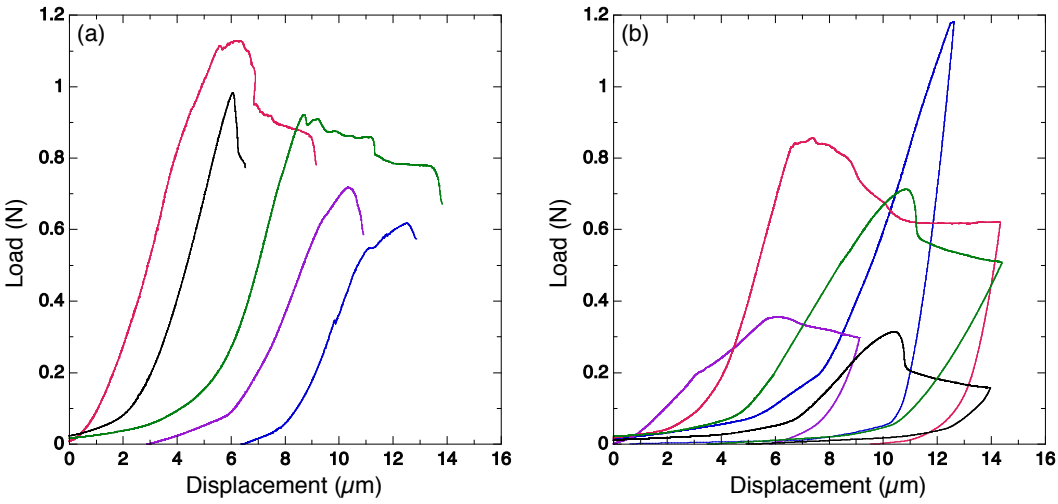

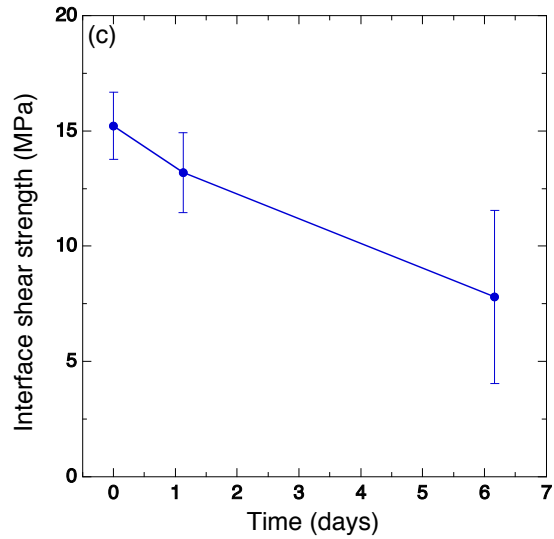

**Fig. 3.** (a) Load-displacement curves of the push-out tests in Mg/PLA composite after 27 hours immersed in PBS. (b) *Idem* after 148 hours. (c) Influence of immersion in PBS on the interface shear strength in Mg/PLA composites.

## 3.3 Degradation micromechanisms

The average value (and the corresponding standard deviation) of the interface shear strength according to eq. (1) is plotted in Fig. 3c as a function of the immersion time in PBS. The reduction in shear strength is very fast and is close to 50% of the initial value after just 6 days.

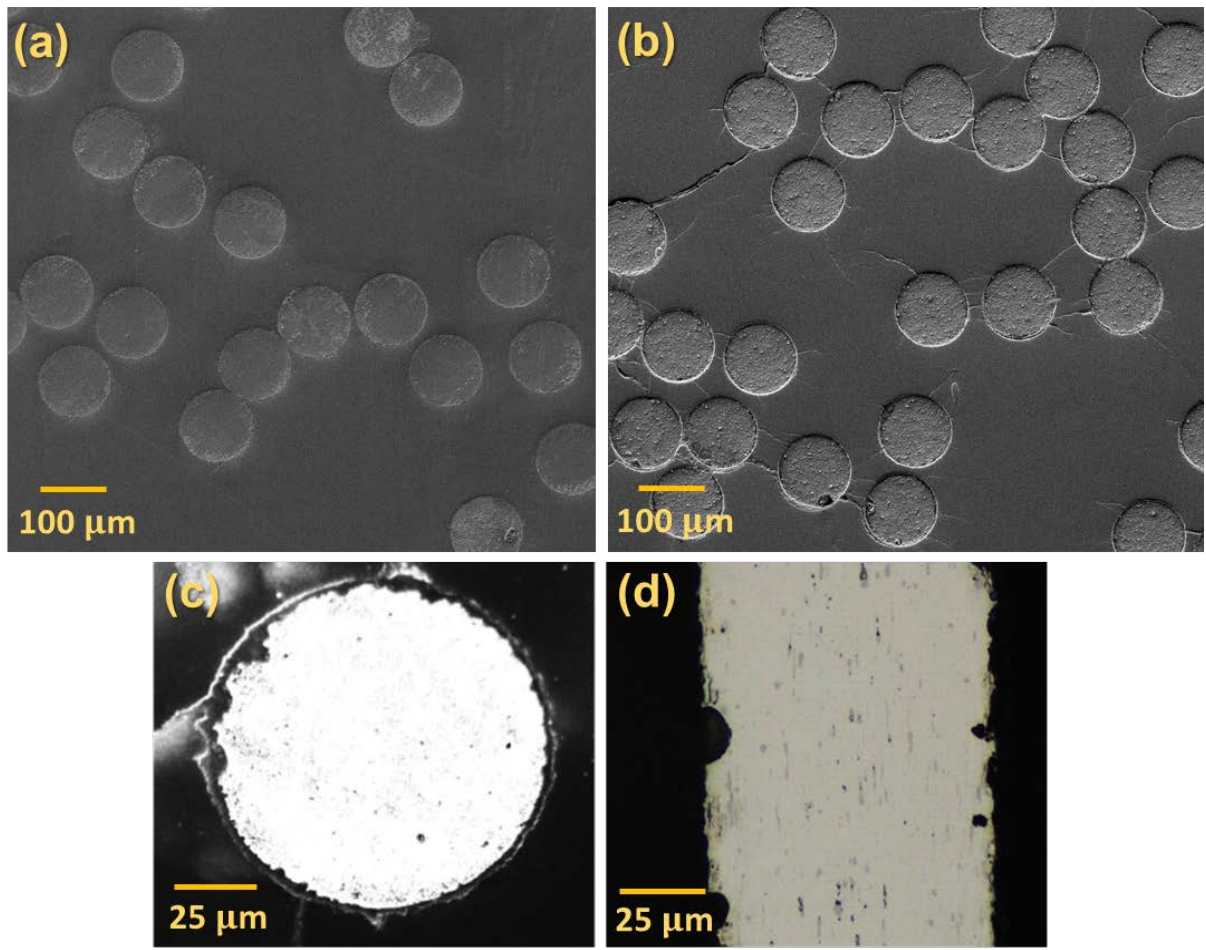

**Fig. 4.** Microstructure of the Mg/PLA composite (a) before immersion and (b) after 148 hours of immersion in PBS showing fiber pitting and matrix cracking. (c) Optical micrograph of the cross-section of a degraded fiber. (c) Optical micrograph of the longitudinal section of a corroded fiber.

The cross-sections of the composite before (Fig. 4a) and after (Fig. 4b) immersion in PBS during 148 h at 37ºC are shown in the SEM image in Figs. 4a and 4b, respectively. Damage is localized at the fiber/matrix interface in the form of corrosion pits in the Mg fibers, which can be seen in the fiber cross-sections perpendicular (Fig. 4c) and parallel (Fig. 4d) to the fiber axis, as well as matrix cracks that tend to link adjacent fibers. The map of the distribution of different chemical elements in the composite cross-section, obtained by energy-dispersive X-ray spectroscopy, is shown in Fig. 5. An oxygen-rich corrosion layer is found at the fiber/matrix interface and sometimes also within the matrix cracks that link the fibers. Traces of Y can also be found in this layer.

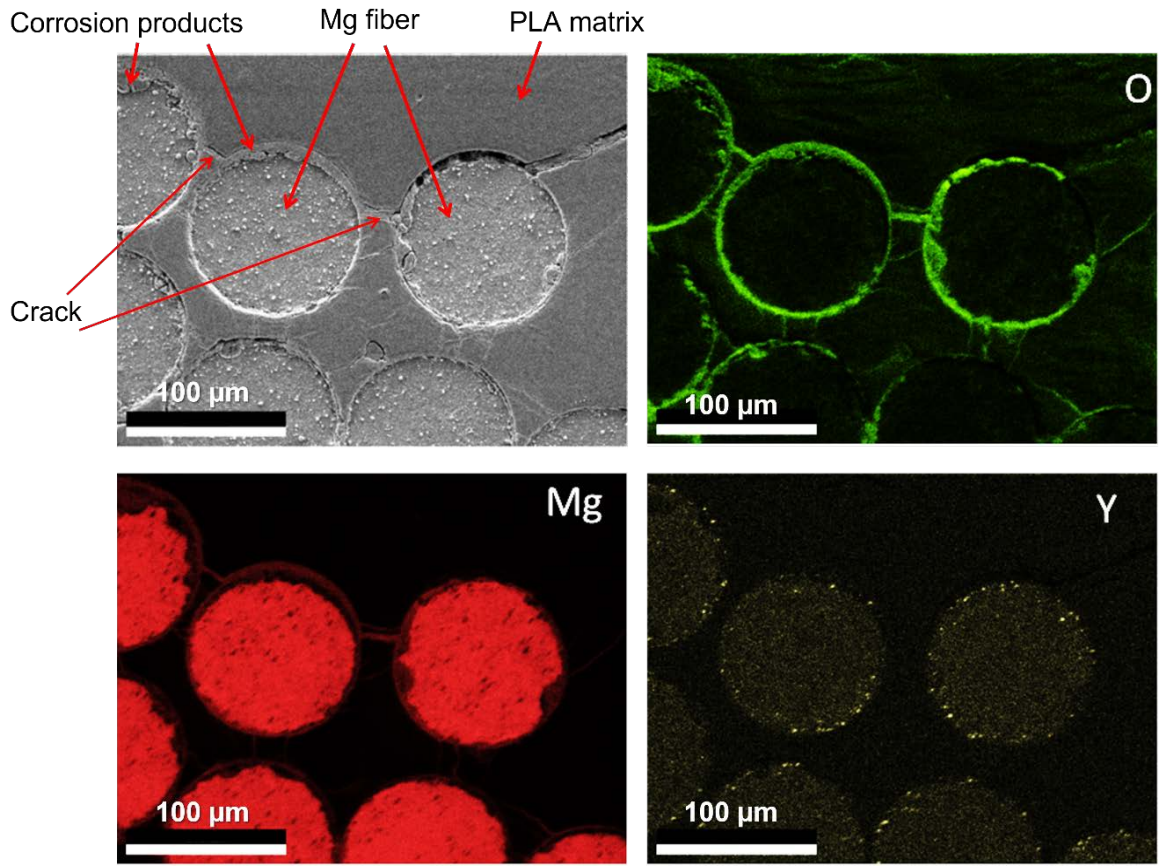

**Fig. 5.** (a) Secondary electron SEM micrograph showing the details of corrosion at the fiber/matrix interface after 148 h in PBS at 37ºC. (b-d) Element map of O, Mg and Y obtained by energy-dispersive X-ray spectroscopy.

**3.4 Cell-composite interaction**

Cells were deposited on the longitudinal and transversal surfaces on the Mg/PLA composites. They were in contact with the PLA in the longitudinal surface and with Mg and PLA in the transversal one. The cells deposited on the longitudinal surface of the composite grew uniformly on the PLA, as it can be seen in SEM micrographs at different magnifications (Figs. 6a and 6b). The lines on the surface of the longitudinal samples are a result of the cutting process (Fig. 6a) and the cells align and respond to this pattern at the microscale. Nevertheless, cells were able to grow homogeneously on the longitudinal surface of the samples made up by PLA. Moreover, they showed a round shape and close cell-cell and cell-material contact due to filipodia structures (Fig. 6c), which indicates a strong attachment of the cells to the surface. A reduction in the number of cells was observed, however, in the transversal cross-section (Fig. 7a).

Moreover, the cells were attached to the PLA but not to the Mg fibers, which showed evidence of the build-up of corrosion products thwarting adhesion (Fig. 7b).

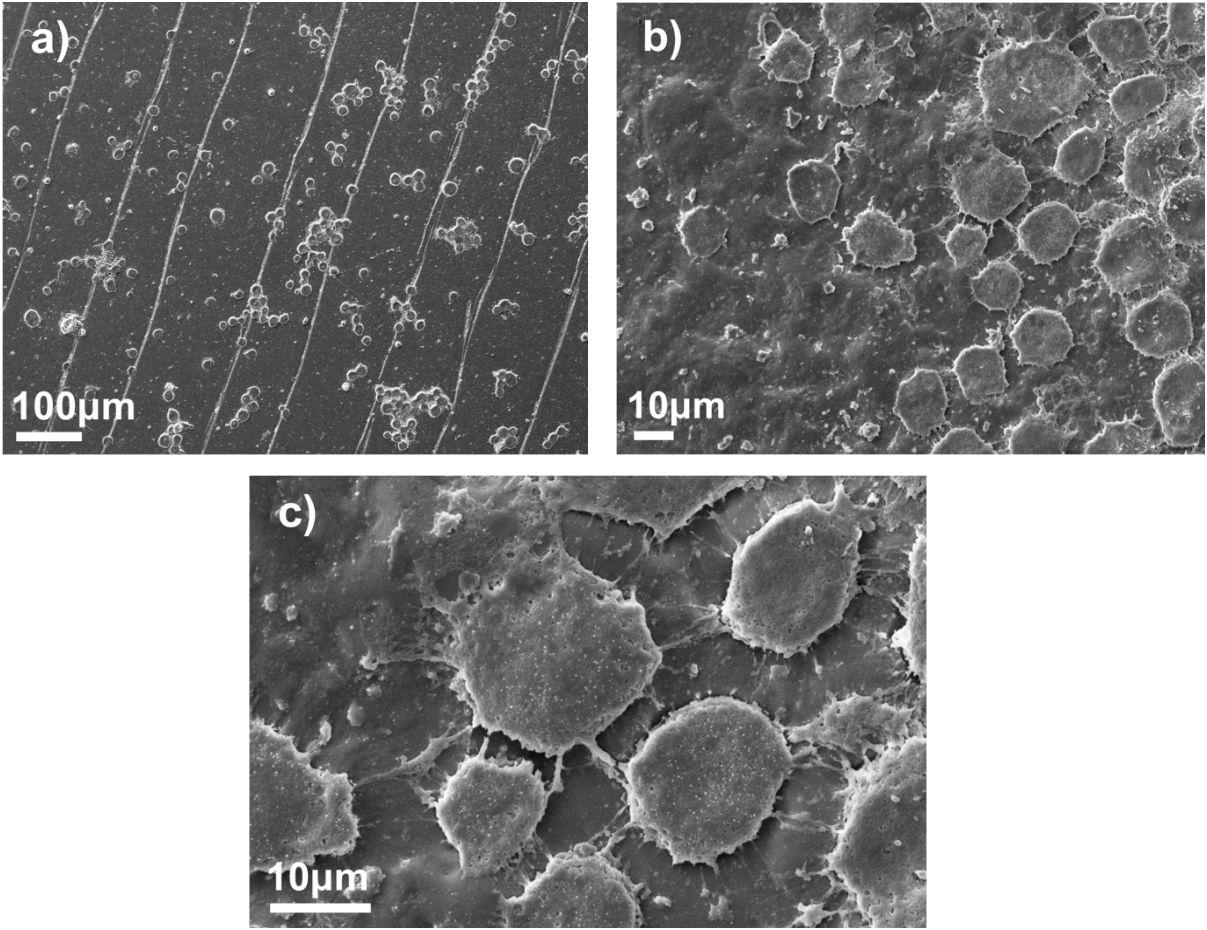

**Fig. 6.** SEM micrographs of the MC3T3-E1pre-osteoblasts deposited on the longitudinal surface of the composite (a,b). Cells attached to PLA showed a round shape and close cell-cell and cell-material contact due to filipodia structures (c).

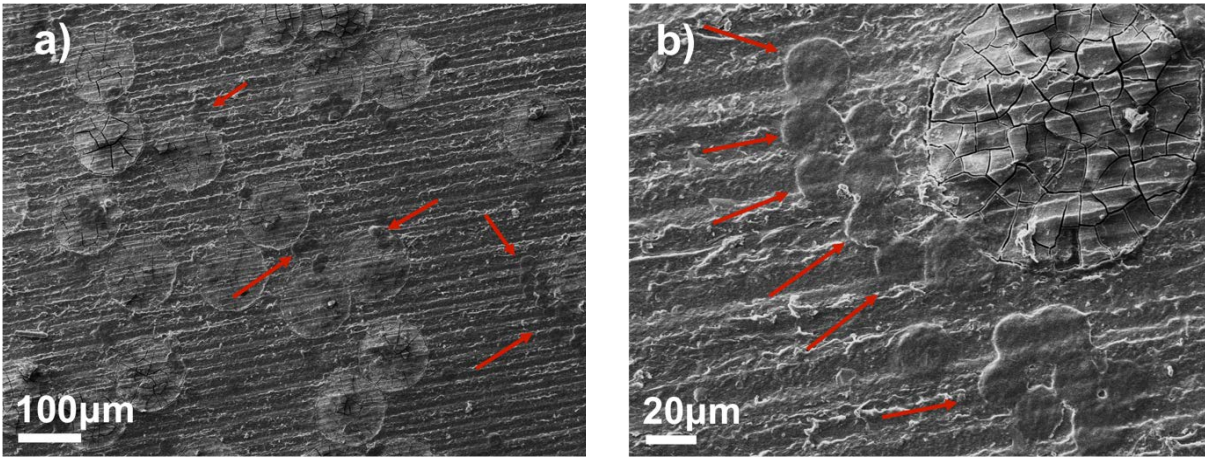

**Fig. 7.** (a, b) SEM micrographs of MC3T3-E1pre-osteoblasts deposited on the transversal surface of the composite. Cells are indicated with red arrows, while fibers show a circular shape with 100 μm in diameter. Cells were mainly attached to the PLA but not to the Mg fibers.

## 4. Discussion

The results presented above show that the diffusion of water from the composite surface towards the Mg fibers took place very quickly when the composite samples were immersed in PBS solution and degradation of the interface strength was already noticeable after 1 day. The interface strength was reduced from 15.2 ± 1.4 MPa to 7.8 ± 3.7 MPa (Fig. 3c) and, more importantly, the radial cracks emerging from the Mg fiber covered the PLA matrix (Fig. 4a). Cracking enabled the rapid diffusion of the water into the Mg fibers accelerating the corrosion and the degradation of the composite material. Mechanical tests of PLA/Mg composites after immersion in corrosive enviroments reported cracks in the PLA matrix [19] and a reduction in the failure strain of composite decreased with degradation time [15].

The degradation of the Mg/PLA composites can be understood from the widely accepted corrosion-induced cracking model of steel reinforced concrete [29]. Corrosion begins at surface of Mg fibers in the first stage by water molecules, leading to a protective layer mainly made up of $Mg(OH)_2$ that contains small amounts of MgO, $Y(OH)_3$ or $Y_2O_3$ [30]. This process was confirmed by the SEM micrographs and the presence of Mg, O and Y at the interface, see Fig. 5. The hydrogen gas ($H_2$) resulting from the reaction diffuses through the PLA matrix. The corrosion layer grows inwards and outwards at same time and severe pitting starts on random locations of the fiber surface by diffusion of chlorine ions which convert $Mg(OH)_2$ into soluble $MgCl_2$ releasing hydroxyl ($OH^{-1}$) ions. The $OH^{-1}$ ions either diffuse out of composite into the medium or react with available carboxylic end groups produced by hydrolytic degradation of PLA, leading to the formation of water [15]. Pitting corrosion reduces severely the strength of the Mg fibers and, thus, the overall load bearing capability of the Mg/PLA composite decreases.

It is known that the corrosion resistance of Mg WE43 alloy is higher than that of Mg and that pitting corrosion happens when the alloy is directly exposed to the corrosive medium [30-31]. Thus, it was surprising that the PLA matrix could not stop pitting corrosion of the Mg fibers, as shown in Figs. 4 and 5. The mechanisms of pitting corrosion could be explained by the cracking of the PLA matrix as a result of the increase in volume of the fibers by the corrosion products. Assuming that the uniform corrosion of the fiber leads to the formation of a layer of thickness $e$ around the fiber surface, the expansion of the fiber induces a biaxial stress state in the PLA matrix at the interface of the fiber, characterized by a compressive stress $\sigma_r = -p$ in the radial direction and a tensile stress $\sigma_\theta = p$ in the circumferential direction. The magnitude of this stresses is given by

$$p = \frac{\bar{E}_{Mg}\bar{E}_{PLA}}{\bar{E}_{PLA}+\bar{E}_{Mg}}\left[(\alpha_{Mg} - \alpha_{PLA})\Delta T + \frac{e}{R}\left(\frac{\beta}{1+\beta}\right)\right] \qquad (2)$$

where $\bar{E}_{PLA} = E_{PLA}/(1 + \nu_{PLA})$ and $\bar{E}_{Mg} = E_{Mg}/(1 - \nu_{Mg} - 2\nu_{Mg}^2)$ and $E_{PLA}$, $\nu_{PLA}$, $E_{Mg}$, $\nu_{Mg}$ stand for the elastic modulus and Poisson's ratio of PLA and Mg. $\alpha_{Mg}$ and $\alpha_{PLA}$ stand for the respective coefficients of thermal expansion and $\Delta T = -148$ °C is the difference between the test temperature (37°C) and the processing temperature (185 °C). R is the fiber radius (50 µm) and $\beta = 0.262$ is the coefficient of linear expansion of corroded region around the fiber due to the transformation of Mg into Mg(OH)$_2$ [32]. The mathematical analysis to obtain expression (2) and the magnitude of $\beta$ are detailed in the appendix while the thermo-elastic constants of Mg and PLA can be found in Table 1.

It should be noted that three difference sources of residual stresses were initially considered in the model to predict the circumferential tensile stress in the PLA matrix at the interface (eq. 2). They were the thermal stresses that arise during the cooling stage after the composite forming process, the swelling of matrix due to the water uptake and the fiber expansion due to the formation of the corrosion layer. Nevertheless, the effect of swelling was not included in the model because it would release the tensile stresses in the matrix and there were not accurate experimental data to the determine the volume expansion due to water uptake. Moreover, the water uptake stops very quickly while cracks evolved with time.

Table 1. Thermo-elastic constants of PLA [33, 34] and Mg [6] and tensile strength of PLA [33].

| $E_{PLA}$ (GPa) | $\nu_{PLA}$ | $\alpha_{PLA}$ (°C$^{-1}$ x 10$^{-6}$) | $\sigma_{PLA}$ (MPa) | $E_{Mg}$ (GPa) | $\nu_{Mg}$ | $\alpha_{Mg}$ (°C$^{-1}$ x 10$^{-6}$) |
|---|---|---|---|---|---|---|
| 3.7 | 0.36 | 70 | 57 | 45.1 | 0.3 | 26 |

Assuming the fracture of PLA under multiaxial stresses ($\sigma_r = -p$, $\sigma_\theta = p$, $\sigma_z = 0$) occurs when Von Mises stress reaches the tensile strength of PLA, $\sigma_{PLA}$, the relationship between $p$ and $\sigma_{PLA}$ can be expressed as

$$\sigma_{PLA} = \frac{1}{\sqrt{2}}\sqrt{(\sigma_r - \sigma_z)^2 + (\sigma_r - \sigma_\theta)^2 + (\sigma_\theta - \sigma_z)^2} = \sqrt{3}p \qquad (3)$$

and, according to eq. (2), failure of the PLA at the interface will take place when the corrosion layer reaches ≈ 1.44 μm. Eq (3) takes into account that the compressive stresses in the radial direction at the interface may induce additional stress concentrations in an incipient radial crack emanating from the corrosion layer. The radial cracks in the PLA facilitate the penetration of water to the interface, leading to pitting corrosion of the fibers and to the fast degradation of the composite because the PLA matrix no longer protects the fiber from corrosion. Further fiber degradation leads to the opening of the cracks in the PLA, which are eventually filled with corrosion products, as shown in Fig. 5.

In order to validate the results of the analytical model, a number of small cracks in the PLA, emanating from the Mg fibers, were selected in the cross-section of the sample immersed in PBS during 27 h. The thickness of the corrosion layer on the fiber was measured at these crack locations, as shown in Fig. 8. They were in the range 1.0 to 1.5 μm, in agreement with the predictions of the theoretical model.

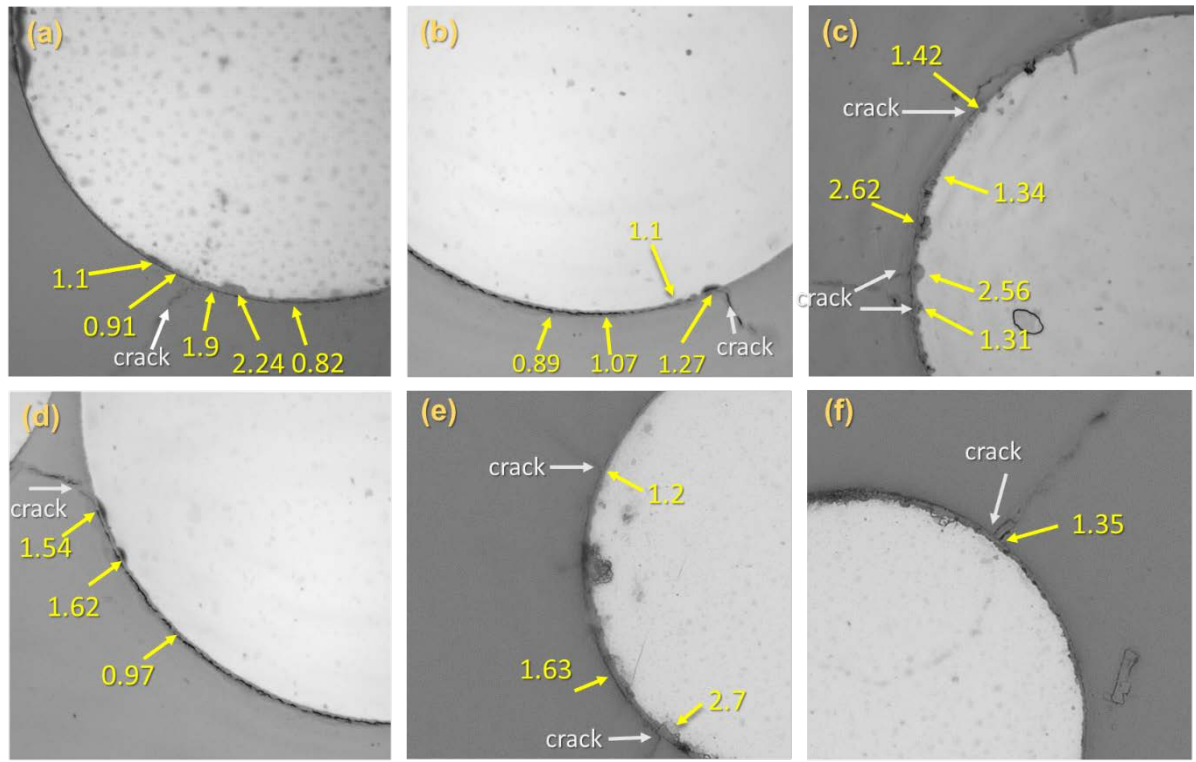

**Fig. 8. (a - f)** Optical micrographs showing small cracks in the PLA in the samples immersed in PBS for 27 h. The thickness of the corrosion layer in the Mg fibers at these locations was measured and the corresponding values (in μm) are indicated in the figure.

The degradation of the Mg fibers also influences the biological response of the composite because viability and metabolic activity of the cells can be affected by abrupt changes in the environment associated with the presence of Mg(OH)$_2$ – that influences the local pH – and release of H$_2$ gas. The evaluation of the cells deposited on the longitudinal and transversal surfaces of the composite showed groups of cells on the PLA but not on the Mg. Moreover, the fraction of cells was much higher in the longitudinal surfaces, where no Mg fibers were present. Thus, it seems that the rapid degradation of the Mg was negative from the viewpoint of biocompatibility, as it has been shown in other investigations [9]. Thus, surface modification of the Mg fibers to enhance the corrosion resistance seems to be a critical step to guarantee the mechanical properties and the biocompatibility of Mg/PLA composites for biomedical applications.

## 5. Conclusions

The shear strength and corrosion resistance of interfaces in poly-lactic acid/Mg fiber composites were studied by means of push-out tests in samples immersed in simulated body fluid. It was found that the interface shear strength of the PLA/Mg interfaces dropped from $15.2 \pm 1.4$ to $7.8 \pm 3.7$ MPa after the sample was immersed in simulated body fluid during 148 hours. The degradation was accompanied by the corrosion of the Mg fibers and the development of radial cracks in the PLA emerging from the corroded fibers due to the expansivity of the $Mg(OH)_2$ formed during corrosion. The cracks facilitated the ingress of water and promoted the formation of corrosion pits in the fibers, decreasing the mechanical properties of the fibers and opening further the radial cracks in the PLA. Biological evaluations tests showed that the early degradation of the Mg fibers did not facilitate the proliferation of pre-osteoblasts cells near the Mg fibers. This phenomenon is attributed to the local changes in the environment produced by the fiber corrosion and/or the corrosion products deposited on the fiber surface. Thus, surface modification of Mg fibers to enhance the corrosion resistance is necessary for further development of Mg/PLA composites for biomedical applications.

**Acknowledgements**

This investigation was supported by the European Union's Horizon 2020 research and innovation programme under the European Training Network BioImpant (Development of improved bioresorbable materials for orthopaedic and vascular implant applications), Marie Skłodowska-Curie grant agreement No 813869. MER acknowledges the support of the Spanish Ministry of Science and Innovation through the Juan de la Cierva-Formation fellowship FJC2019-039925-I

**Appendix.** Analytical model for the stresses in the matrix at the interface due to the uniform corrosion of the fibers.

A simplified thermo-elastic analysis is presented to determine the stresses in the PLA matrix at the interface with a circular Mg fiber as a result of the thermal expansion mismatch between matrix and fibers and of the expansion of the corrosion products in an annular region of the fiber of thickness $e_0$. The circular fiber of radius $R$ is embedded in an infinite matrix and the thermo-elastic problems of a cylindrical cavity subjected to an uniform internal pressure $p$ (Fig. A1a) and of a cylindrical fiber subjected to an uniform lateral pressure $p$ (Fig. A1b) are solved independently.

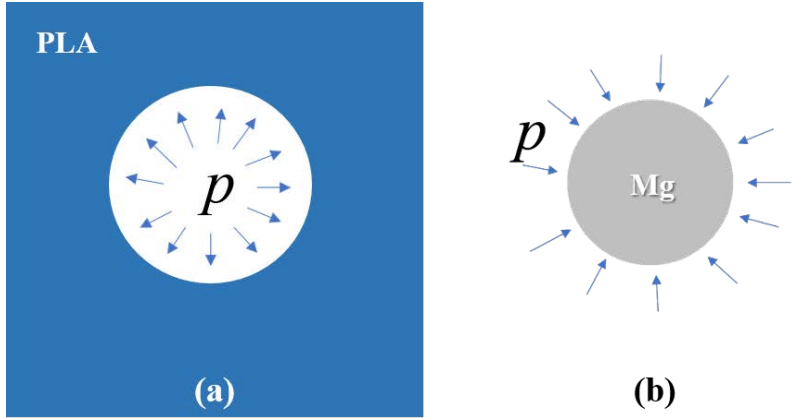

**Fig. A1.** Schematics of the thermo-elastic problems. (a) Cylindrical cavity subjected to an uniform internal pressure -$p$. (b) Cylindrical fiber subjected to an uniform lateral pressure -$p$.

The elastic solution of the cylindrical cavity subjected to an internal pressure $P$ can be found in the literature [1]. The radial ($\sigma_r$) and circumferential ($\sigma_\theta$) stresses in the matrix decrease with $R^{-2}$ and they are maximum at r=R

$$\sigma_r(r = R) = -p, \quad \sigma_\theta(r = R) = p, \quad \sigma_z(r = R) = \nu_{PLA}(\sigma_r + \sigma_\theta) = 0 \quad (A1)$$

where $\nu_{PLA}$ stands for the Poisson's ratio of the PLA matrix. The circumferential strain at the interface, including the contribution of thermal strains associated with the temperature change from the processing temperature up to the test temperature, $\Delta T$, is given by

$$\varepsilon_\theta^{PLA} = \frac{p}{E_{PLA}} - \frac{\nu_{PLA}}{E_{PLA}}(-p) + \alpha_{PLA}\Delta T = \frac{p}{\bar{E}_{PLA}} + \alpha_{PLA}\Delta T \quad \text{where} \quad \bar{E}_{PLA} = \frac{E_{PLA}}{1+\nu_{PLA}} \quad (A3)$$

where $E_{PLA}$ and $\alpha_{PLA}$ stand for the elastic modulus and linear thermal expansion coefficient of PLA.

The stress state of the circular fiber subjected to a lateral pressure $p$ is constant and -under plane strain conditions- is given by

$$\sigma_r = -p, \quad \sigma_\theta = -p, \quad \sigma_z = \nu_{Mg}(\sigma_r + \sigma_\theta) = -2\nu_{Mg}p \tag{A2}$$

where $\nu_{Mg}$ stands for the Poisson's ratio of the Mg fiber. The circumferential strain in the fiber -including thermal strains- is given by

$$\varepsilon_\theta^{Mg} = \frac{-p}{E_{Mg}} - \frac{\nu_{Mg}}{E_{Mg}}(-p - 2\nu_{Mg}p) + \alpha_{Mg}\Delta T = -\frac{p}{\bar{E}_{Mg}} + \alpha_{Mg}\Delta T \text{ where } \bar{E}_{Mg} = \frac{E_{Mg}}{1-\nu_{Mg}-2\nu_{Mg}^2} \tag{A3}$$

where $E_{Mg}$ and $\alpha_{Mg}$ stand for the elastic modulus and linear thermal expansion coefficient of Mg.

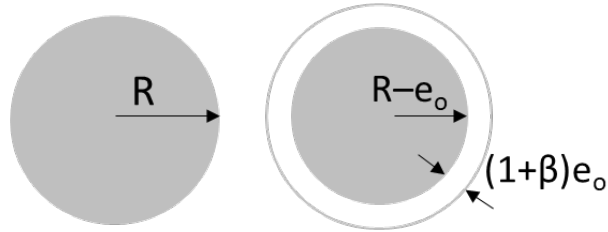

**Fig. A2.** Schematic of the change in the fiber radius due to the transformation of a layer of Mg of thickness $e_0$ in Mg(OH)$_2$.

During the uniform corrosion of the fiber, a layer of Mg of thickness $e_0$ is transformed in Mg(OH)$_2$, but the volume expansion associated with corrosion leads to a layer of corroded material of thickness $e = (1 + \beta)e_0$ (Fig. A2). $\beta$ stands for the linear expansion coefficient due to the formation of the corrosion products. It can be determined as one third of the total increase in volume when one mol of Mg ($m_{Mg} = 24.3$ g) is transformed in one mol of Mg(OH)$_2$ ($m_{Mg(OH)_2} = 58.42$ g) as

$$\beta = \frac{1}{3}\left(\frac{\frac{m_{Mg(OH)_2}}{\rho_{Mg(OH)_2}} - \frac{m_{Mg}}{\rho_{Mg}}}{\frac{m_{Mg}}{\rho_{Mg}}}\right) = 0.262 \tag{A4}$$

where $\rho_{Mg} = 1.738\ Kg/m^3$ = and $\rho_{Mg(OH)_2} = 2.34\ Kg/m^3$ stand for the densities of Mg and Mg(OH)$_2$.

The expansion in the fiber radius leads to an additional contribution to the circumferential strain, $\varepsilon_\theta^{corr}$, which is given by,

$$\varepsilon_\theta^{corr} = \frac{2\pi(R-e_0+(1+\beta)e_0)-2\pi R}{2\pi R} = \frac{e}{R}\left(\frac{\beta}{1+\beta}\right) \tag{A5}$$

As the thickness of this corrosion layer is very small, it can be assumed that the circumferential strains at the interface between the between the PLA matrix and the corrosion layer are equal, leading to the compatibility equation

$$-\frac{p}{\bar{E}_{Mg}} + \alpha_{Mg}\Delta T + \varepsilon_\theta^{corr} = \frac{p}{\bar{E}_{PLA}} + \alpha_{PLA}\Delta T \tag{A6}$$

that can be solved to obtain the pressure $p$ as a function of the thickness of the corrosion layer $e$,

$$p = \frac{\bar{E}_{Mg}\bar{E}_{PLA}}{\bar{E}_{PLA}+\bar{E}_{Mg}}\left[(\alpha_{Mg}-\alpha_{PLA})\Delta T + \frac{e}{R}\left(\frac{\beta}{1+\beta}\right)\right] \tag{A7}$$

[1] E. Torroja, Elasticidad, Dossat, Madrid, 1967.